\documentclass[12pt]{article}
\usepackage{amssymb}
\usepackage[dvips]{graphicx,psfrag}
\usepackage{array}
\usepackage{epsfig}

\makeatletter
\def\alt{\mathrel{\mathpalette\gl@align<}}
\def\agt{\mathrel{\mathpalette\gl@align>}}
\def\gl@align#1#2{\lower.6ex\vbox{\baselineskip\z@skip\lineskip\z@
\ialign{$\m@th#1\hfil##\hfil$\crcr#2\crcr\sim\crcr}}} \makeatother

\def\bwt{\begin{widetext}}
\def\ewt{\end{widetext}}
\def\be{\begin{equation}}
\def\ee{\end{equation}}
\def\bea{\begin{eqnarray}}
\def\eea{\end{eqnarray}}
\def\bean{\begin{eqnarray*}}
\def\eean{\end{eqnarray*}}
\def\bary{\begin{array}}
\def\eary{\end{array}}
\def\bit{\begin{itemize}}
\def\eit{\end{itemize}}

\def\su5u1{SU(5) \times U(1)}
\def\fsu5u1{SU(5) \times U(1)'}
\def\so10{SO(10)}
\def\sq20{SO(10) \times SO(10)}

\begin{document}
\begin{flushright}
{\tt hep-ph/0604217}\\
BA-06-13\\
\end{flushright}
\vspace*{1.0cm}
\begin{center}
{\baselineskip 25pt \Large{\bf
Higgs Boson Mass from Orbifold GUTs \\ with Split Supersymmetry \\[2.5mm]
}}

\vspace{1cm}

{\large Ilia Gogoladze$^a$\footnote {On a leave of absence from:
Andronikashvili Institute of Physics, GAS, 380077 Tbilisi, Georgia.
\\ \hspace*{0.5cm} }, Tianjun Li$^{b,c}$
,
V. N. {\c S}eno$\breve{\textrm{g}}$uz$^d$
  and Qaisar
Shafi$^d$ }
\vspace{.5cm}

{\small {\it $^a$Department of Physics and Astronomy, University of
Delaware, Newark, DE 19716, USA \\
$^b$Department of Physics and Astronomy, Rutgers University,
Piscataway, NJ 08854, USA\\
$^c$Institute of Theoretical Physics, Chinese Academy of Sciences,
 Beijing 100080, P. R. China \\
$^d$ Bartol Research Institute, Department of Physics and Astronomy,
University of Delaware, Newark, DE 19716, USA }}



\vspace{1.0cm} {\bf Abstract}

\end{center}

\baselineskip 16pt

We consider orbifold GUTs with ${\cal N}=1$ supersymmetry in which the Standard
Model (SM) gauge couplings are unified at $M_{\rm GUT} \simeq 2\times 10^{16}$ GeV with one
of the third family (charged) Yukawa couplings. With split supersymmetry
the SM Higgs mass is estimated to be $131\pm10$ GeV for gauge-top quark
Yukawa unification, which increases to $146\pm 8$ GeV for gauge-bottom quark
(or gauge-tau lepton) Yukawa unification.

\thispagestyle{empty}

\bigskip
\newpage

\addtocounter{page}{-1}

\section{Introduction}
In a recent paper \cite{Gogoladze:2006ps} the Higgs boson mass of the Standard
Model (SM) was estimated within
the framework of 7-dimensional orbifold grand unified
theories (7D orbifold GUTs) with ${\cal N}=1$ supersymmetry, compactified
on the orbifold $M^4\times T^2/Z_6 \times S^1/Z_2$ (for details see Appendix A and
Refs. \cite{Gogoladze:2006ps,Barger:2005gn,GY,Li:2001tx}). It was assumed in \cite{Gogoladze:2006ps} 
that the 4D ${\cal N}=1$ supersymmetry (SUSY) is
broken at $M_{\rm GUT}$ and, depending on the gauge--Yukawa unification assumed,
the Higgs mass varies between $135\pm6$ GeV and $144\pm 4$ GeV. The unification condition
fixes $\tan\beta$ (ratio of the two Minimal Supersymmetric Standard
Model (MSSM) VEVs) which makes the Higgs mass
prediction possible.

In this paper we extend the discussion in \cite{Gogoladze:2006ps} to the case of
split supersymmetry in which the supersymmetric scalars are at an intermediate
scale $m_S$ ($\sim10^6$--$10^{11}$ GeV), while gauge coupling unification is
achieved with TeV-scale gauginos and Higgsinos. As in the MSSM, the Higgs
mass is determined by its quartic coupling $\lambda$ generated by the
supersymmetric D-term at $m_S$ and the one-loop top quark Yukawa corrections
to the Higgs potential at the weak scale. Unlike the MSSM where the Higgs mass
prediction also depends on the soft supersymmetry breaking A terms and
stop masses due to radiative corrections, in split supersymmetry the A terms
are suppressed relative to $m_S$, and the threshold corrections from the
stops are tiny. As a result, the Higgs boson mass can be calculated
quite reliably, although it depends on the unknown parameter $m_S$.

The plan of this paper is as follows. In Section 2 we summarize the 
7D $SU(7)$ orbifold models and show how gauge--Yukawa 
coupling unification can be realized. Section 3 is a brief review
of split SUSY. Section 4 is devoted to the Higgs boson mass predictions.
We conclude in Section 5.

\section{$SU(7)$ Orbifold Models}

To realize the unification of gauge couplings and one of the Yukawa
couplings for the third-family quarks and lepton at the GUT scale,   we consider
a 7D ${\cal N}=1$ supersymmetric $SU(7)$ gauge theory compactified
on the orbifold $M^4\times T^2/Z_6 \times S^1/Z_2$ (for some details see
Appendix A). We find that $SU(7)$ is the smallest gauge group which allows us to implement
these gauge--Yukawa coupling unification conditions.

The ${\cal N}=1$ supersymmetry in 7D has 16 supercharges corresponding
to ${\cal N}=4$ supersymmetry in 4-dimension (4D), and only the
gauge supermultiplet can be introduced in the bulk.  This multiplet
can be decomposed under  4D
 ${\cal N}=1$ supersymmetry into a gauge vector
multiplet $V$ and three chiral multiplets $\Sigma_1$, $\Sigma_2$,
and $\Sigma_3$,  all in the adjoint representation, where the fifth
and sixth components of the gauge field, $A_5$ and $A_6$, are
contained in the lowest component of $\Sigma_1$, and the seventh
component of the gauge field $A_7$ is contained in the lowest
component of $\Sigma_2$. As pointed out in Ref.~\cite{NMASWS}
the bulk action in the Wess-Zumino gauge and in 4D ${\cal N}=1$
supersymmetry notation contains trilinear terms involving  the
chiral  multiplets $\Sigma_i$. Appropriate choice of the orbifold
enables us to identify some of them as the SM Yukawa couplings~\cite{GY}.

To break the $SU(7)$ gauge symmetry, we select the following
$7\times 7$ matrix representations for $R_{\Gamma_T}$ and
$R_{\Gamma_S}$ defined in Appendix A
\begin{eqnarray}\label{bb1}
R_{\Gamma_T} &=& {\rm diag} \left(+1, +1, +1,
 \omega^{n_1}, \omega^{n_1}, \omega^{n_1}, \omega^{n_2} \right),
 \label{bb3}
\end{eqnarray}
\begin{eqnarray}\label{bb4}
 R_{\Gamma_S} &=& {\rm diag} \left(+1, +1, +1, +1, +1,
-1, -1 \right),
\end{eqnarray}
where $n_1$ and $n_2$ are positive integers, and $n_1 \not= n_2$.
 Then, we obtain
\begin{eqnarray}
&& \{SU(7)/R_{\Gamma_T}\} ~=~ SU(3)_C\times SU(3)\times U(1) \times
U(1)^{\prime},
\nonumber \\
 &&\{SU(7)/R_{\Gamma_S}\} ~=~ SU(5)\times SU(2) \times
U(1),
\end{eqnarray}
\begin{eqnarray}
 \{SU(7)/\{R_{\Gamma_T} \cup R_{\Gamma_S}\}\}
~=~ SU(3)_C\times SU(2)_L\times U(1)_Y \times U(1)_{\alpha} \times
U(1)_{\beta}. \label{bb2}
\end{eqnarray}
So, the 7D ${\cal N}= 1 $
supersymmetric gauge symmetry $SU(7)$ is broken down to  4D
${\cal N}=1$ supersymmetric gauge symmetry $SU(3)_C\times
SU(2)_L\times U(1)_Y \times U(1)_{\alpha} \times U(1)_{\beta} $~\cite{Li:2001tx}.
In Eq. (\ref{bb2}) we see the appearance of two U(1) gauge
symmetries which we assume can be spontaneously broken at or close
to $M_{\rm GUT}$ by the usual Higgs mechanism. It is conceivable
that these two symmetries can play some useful role as flavor
symmetries, but we will not pursue this any further here.
A judicious choice of $n_1$ and $n_2$ will enable us to obtain the
desired zero modes from the chiral multiplets  $\Sigma_i$ defined in
Appendix A.

The $SU(7)$ adjoint representation $\mathbf{48}$ is decomposed under
the $SU(3)_C\times SU(2)_L\times U(1)_Y \times U(1)_{\alpha} \times
U(1)_{\beta}$ gauge symmetry as:
\begin{equation}
\mathbf{48} = \left(
\begin{array}{cccc}
\mathbf{(8,1)}_{Q00} & \mathbf{(3, \bar 2)}_{Q12}
& \mathbf{(3, 1)}_{Q13} & \mathbf{(3,1)}_{Q14} \\
 \mathbf{(\bar 3,  2)}_{Q21} & \mathbf{(1,3)}_{Q00}
& \mathbf{(1, 2)}_{Q23} & \mathbf{(1, 2)}_{Q24} \\
\mathbf{(\bar 3, 1)}_{Q31} & \mathbf{(1, \bar 2)}_{Q32}
& \mathbf{(1, 1)}_{Q00} & \mathbf{(1, 1)}_{Q34} \\
\mathbf{(\bar 3, 1)}_{Q41} & \mathbf{(1, \bar 2)}_{Q42}
& \mathbf{(1, 1)}_{Q43} & \mathbf{(1, 1)}_{Q00}
\end{array}
\right) +  \mathbf{(1,1)}_{Q_{00}}\, ,
\label{48arj}\end{equation} where the  $\mathbf{(1,1)}_{Q00}$ in the
third and fourth diagonal entries of the matrix and the last term
$\mathbf{(1,1)}_{Q_{00}}$ denote the gauge fields
associated with $U(1)_Y \times U(1)_{\alpha} \times U(1)_{\beta} $.
The subscripts $Qij$, which are anti-symmetric ($Qij=-Qji$), are the
charges under  $U(1)_Y \times U(1)_{\alpha} \times U(1)_{\beta}$.
The subscript $Q00~=~(\mathbf{0}, \mathbf{0}, \mathbf{0})$, and the
other subscripts $Qij$ with $i\not= j$ will be given for each model
explicitly.

\subsection{Unification of Gauge and Top Quark Yukawa Coupling }

To achieve gauge and top quark Yukawa coupling unification at
$M_{\rm GUT}$, we make the following choice
\begin{eqnarray}
n_1~=~5~~ {\rm and}~~n_2~=~2~~~{\rm or}~~~3\,,~\, \label{N-numberA}
\end{eqnarray}
in Eq. (\ref{bb1}).  This allows  us to obtain zero modes from
$\Sigma_i$ corresponding to a pair of Higgs doublets $H_u$ and
$H_d$, as well as the left- and right-handed top quark superfields.

\renewcommand{\arraystretch}{1.4}
\begin{table}[ht]
\begin{center}
\begin{tabular}{|c|c|}
\hline
Chiral Fields & Zero Modes  \\
\hline\hline
$\Sigma_1$ & $Q_3$:~ $\mathbf{(3, \bar 2)}_{Q12}$ \\
\hline $\Sigma_2$ & ~$H_u$:~ $\mathbf{(1, 2)}_{Q23}$;
~$H_d$:~ $\mathbf{(1, \bar 2)}_{Q32}$ \\
\hline
$\Sigma_3$ & $t^c$: $\mathbf{(\bar 3, 1)}_{Q31}$ \\
\hline
\end{tabular}
\end{center}
\vspace{-0.3cm} \caption{\small Zero modes from  the chiral
multiplets $\Sigma_1$, $\Sigma_2$ and $\Sigma_3$  with gauge and top
quark Yukawa coupling unification. \label{Spectrum-UP-7D}}
\end{table}

 The generators for the gauge symmetry $U(1)_Y
\times U(1)_{\alpha} \times U(1)_{\beta}$ are as follows:
\begin{eqnarray}
&&T_{U(1)_{Y}} \equiv {1\over 6} {\rm diag}\left(1, 1, 1, 0, 0, -3,
0 \right) + ~{{\sqrt {21}}\over {42}} {\rm diag}\left(1, 1, 1, 1, 1,
1, -6 \right), \label{SU7-GU1Y} \nonumber \\ && T_{U(1)_{\alpha}}
\equiv -{{\sqrt {21}}\over {2}} ~{\rm diag}\left(1, 1, 1, 0, 0, -3,
0 \right) + ~{\rm diag}\left(1, 1, 1, 1, 1, 1, -6 \right), \nonumber
\\&& T_{U(1)_{\beta}} \equiv {\rm diag}\left(1, 1, 1, -2, -2, 1, 0
\right), \label{SU7-GU1A}
\end{eqnarray}
With a canonical normalization ${\rm tr}[T_i^2]=1/2$ of non-abelian
generators, from Eq. (\ref{SU7-GU1A}) we find ${\rm tr}
[T_{U(1)_{Y}}^2]=5/6$. We denote the $U(1)_Y$, $SU(2)_L$, and
$SU(3)_C$ gauge couplings as $g_Y$, $g_2$, and $g_3$, respectively.
So, for  $k_Y g_Y^2 = g_2^2 = g_3^2$ at the GUT
scale, this gives us the canonical $U(1)_Y$ normalization
$k_Y = 5/3$.

The charge assignments $Qij$ from Eq. (\ref{48arj}) are as follows:
\begin{eqnarray}
 &&Q12 = \left ( \mathbf{1\over 6}, \mathbf{-{{\sqrt {21}}\over 2}}, \mathbf{3}\right ),~~~~~
 Q14 = \left (\mathbf{{{1+{\sqrt {21}}}\over {6}}}, \mathbf{{{14-{\sqrt {21}}}\over {2}}},
\mathbf{1}\right ),
 \nonumber \\
&& Q13= \left (\mathbf{2\over 3}, \mathbf{-2{\sqrt {21}}},
\mathbf{0} \right ),
 ~~~~~  Q23 =  \left ( \mathbf{{1\over 2}},
\mathbf{-{{3{\sqrt {21}}}\over {2}}}, \mathbf{-3}\right ),
\nonumber \\
&& Q24=\left ( \mathbf{{{\sqrt {21}}\over 6}}, \mathbf{{7}},
\mathbf{-2}\right
 ), ~~~~~~~ Q34=\left
(\mathbf{{{-3+{\sqrt {21}}}\over {6}}}, \mathbf{{14+3{\sqrt
{21}}}\over {2}}, \mathbf{1}\right ).
\end{eqnarray}

Substituting Eq. (\ref{N-numberA}) in Eqs. (\ref{bb3})--(\ref{bb4})
and employing the $Z_6 \times Z_2$ transformation properties Eqs.
(\ref{bb7})--(\ref{bb10}) for the decomposed components of the
chiral multiplets $\Sigma_i$, we obtain the zero modes given in
Table 1. We can identify them as a pair of Higgs doublets as well
as the left- and right-handed top quark superfields, as desired.

From the trilinear term in the 7D bulk action  the top quark Yukawa
coupling is contained in the term
\begin{eqnarray}
\label{bs1}  \int d^7 x \left[ \int d^2 \theta \ g_7 Q_3 t^c H_u
 + h.c.\right],
\end{eqnarray}
where $g_7$ is the $SU(7)$ gauge coupling at the compactification
scale, which  for simplicity,  we identify as  $M_{\rm GUT}$.
 We
will ignore  brane localized gauge kinetic terms, which may be
suppressed by taking  $VM_{*}\gtrsim O(100)$, where $V$ denotes the
volume of the extra dimensions and $M_{*}$  is the cutoff scale
\cite{Orbifold}. With these caveats we obtain the 4D gauge--top
quark Yukawa coupling unification at $M_{\rm GUT}$
\begin{eqnarray}
g_1=g_2=g_3=y_t=g_7/\sqrt{V}, \label{un}
\end{eqnarray}
where  $g_1\equiv {\sqrt {5/3}} g_Y$, and
$y_t$ is the top quark Yukawa coupling.

As far as the remaining SM fermions are concerned, we note that
on the 3-brane at the $Z_6\times Z_2$ fixed point
 $(z, y)= (0, 0)$, the preserved gauge symmetry
is $SU(3)_C\times SU(2)_L\times U(1)_Y \times U(1)_{\alpha}\times
U(1)_{\beta}$. Thus, on the observable 3-brane at $(z, y)= (0, 0)$,
we can introduce  the first two
 families of the SM quarks and leptons,
 the right-handed bottom quark,  the $\tau$ lepton doublet,
and the right-handed $\tau$ lepton. The $U(1)_{\alpha}\times
U(1)_{\beta}$ anomalies can be canceled by assigning  suitable
charges to the SM quarks and leptons.  For example,
 under  $U(1)_{\alpha}\times U(1)_{\beta}$ the  charges for
the first-family quark doublet  and the right-handed up quark can be
respectively $(\mathbf{{\sqrt {21}}/ 2}, \mathbf{-3})$ and
$(\mathbf{-2{\sqrt {21}}}, \mathbf{0})$, while the charges of
remaining SM fermions are zero.

\subsection{Unification of Gauge and Down-Type Yukawa Couplings }

To realize gauge--bottom quark Yukawa coupling unification,
 we make the following  choice  in Eq.
(\ref{bb1}):
\begin{eqnarray}
n_1~=~5~,~~n_2~=~2~{\rm or}~3~.~\, \label{bt1}
\end{eqnarray}
The identification of $U(1)_Y$ differs from the  previous  subsection.
The generators of  $U(1)_Y \times U(1)_{\alpha} \times U(1)_{\beta}$
are defined as follows:
\begin{eqnarray}
&&T_{U(1)_{Y}} \equiv -{1\over 6}{\rm diag}\left(0, 0, 0, 1, 1, -2,
0 \right) ~+ {{\sqrt {7}}\over {21}}{\rm diag}\left(1, 1, 1, 1,
1, 1, -6 \right) ,~\, \label{SU7-GD1Y} \nonumber \\
&&T_{U(1)_{\alpha}} \equiv 2 {\sqrt {7}} ~ {\rm diag}\left(0, 0, 0,
1, 1, -2, 0
\right)+ {\rm diag}\left(1, 1, 1, 1, 1, 1, -6 \right), \nonumber \\
&&T_{U(1)_{\beta}} \equiv {\rm diag}\left(1, 1, 1, -1, -1, -1, 0
\right). \label{SU7-GD1A}
\end{eqnarray}
Note that $k_Y$ is also $5/3$ in this case.

\renewcommand{\arraystretch}{1.5}
\begin{table}[htb]
\begin{center}
\begin{tabular}{|c|c|}
\hline
Chiral Fields & Zero Modes  \\
\hline\hline
$\Sigma_1$ & $Q_3$:~ $\mathbf{(3, \bar 2)}_{Q12}$ \\
\hline $\Sigma_2$ & ~$H_d$:~ $\mathbf{(1, 2)}_{Q23}$;
~$H_u$:~ $\mathbf{(1, \bar 2)}_{Q32}$ \\
\hline
$\Sigma_3$ & $b^c$: $\mathbf{(\bar 3, 1)}_{Q31}$ \\
\hline
\end{tabular}
\end{center}
\vspace{-0.3cm}
 \caption{\small Zero modes from the chiral multiplets
$\Sigma_1$, $\Sigma_2$ and $\Sigma_3$  with gauge  and bottom quark
Yukawa coupling unification. \label{Spectrum-DOWN-7D}}
\end{table}

The corresponding charges $Qij$ are:
\begin{eqnarray}
&&Q12=\left (\mathbf{1\over 6}, \mathbf{-2{\sqrt {7}}},
\mathbf{2}\right ),~~~~~ Q13=\left (\mathbf{-{1\over 3}},
\mathbf{4{\sqrt {7}}}, \mathbf{2}\right ),
\nonumber \\
&&Q14= \left (\mathbf{{{\sqrt {7}}\over {3}}}, \mathbf{7},
\mathbf{1}\right ),~~~~~~~~~ Q34=\left (\mathbf{{1+{\sqrt {7}}}\over
3}, \mathbf{7-4{\sqrt {7}}}, \mathbf{-1}\right),
\nonumber \\
&&Q24 =\left (\mathbf{-{1+2{\sqrt {7}}}\over 6}, \mathbf{7+2{\sqrt
{7}}}, \mathbf{-1}\right ),~~~~~ Q23=\left (\mathbf{-{1\over 2}},
\mathbf{6{\sqrt {7}}}, \mathbf{0}\right ).
\end{eqnarray}

In Table \ref{Spectrum-DOWN-7D}, we present the zero modes from  the
chiral multiplets $\Sigma_1$, $\Sigma_2$ and $\Sigma_3$. We identify
them as the left-handed doublet   ($Q_3$), 
one pair of Higgs doublets $H_u$ and $H_d$, and the 
right-handed bottom quark $b^c$. From
the trilinear term in the 7D bulk action in Eq. (\ref{action7}) we
obtain the bottom quark Yukawa coupling
\begin{eqnarray}
 \int d^7 x \left[ \int d^2 \theta \ g_7 Q_3 b^c H_d
 + h.c. \right].~\,
\end{eqnarray}
Thus, at $M_{\rm GUT}$ we have
\begin{eqnarray}
g_1=g_2=g_3=y_b=g_7/\sqrt{V}, \label{mm11}
\end{eqnarray}
where $y_b$ is the bottom quark Yukawa coupling to $H_d$.

Finally, to realize the gauge--tau lepton Yukawa coupling unification, we set
\begin{eqnarray}
n_1~=~4~,~~n_2~=~3;~{\rm or}~ n_1~=~3~,~~n_2~=~2~.~\,
\label{N-numberB}
\end{eqnarray}
The generators for  $U(1)_Y \times U(1)_{\alpha} \times
U(1)_{\beta}$ are as follows:
\begin{eqnarray}
&&T_{U(1)_{Y}} \equiv  {1\over 2} ~ {\rm diag}\left(0, 0, 0, 0, 0,
1, -1 \right) - ~{{\sqrt {7}}\over {42}} {\rm diag}\left(4, 4, 4,
-3, -3, -3, -3 \right),~ \nonumber \\ && T_{U(1)_{\beta}} \equiv
-{{2 \sqrt {7}}\over {3}} ~{\rm diag}\left(0, 0, 0, 0, 0, 1, -1
\right) -{1\over 3} ~{\rm diag}\left(4, 4, 4, -3, -3, -3, -3
\right), \nonumber \\ &&T_{U(1)_{\alpha}} \equiv {\rm diag}\left(0,
0, 0, 1, 1, -1, -1 \right).
\end{eqnarray}

The $U(1)_Y \times U(1)_{\alpha} \times U(1)_{\beta}$ charges
$Qij$ are
\begin{eqnarray}
 &&Q12=\left (\mathbf{-{{\sqrt {7}}\over 6}}, 
\mathbf{-{{7}\over 3}}, \mathbf{-1}\right),~~~~~
Q13=\left (\mathbf{-{{3+\sqrt {7}}\over 6}},
\mathbf{-{{7-2\sqrt {7}}\over 3}}, \mathbf{1}\right ), \nonumber \\
&&  Q23=\left (\mathbf{-{1\over 2}}, 
\mathbf{{2 \sqrt{7}}\over {3}}, \mathbf{2}\right),~~~~~~~~~~  
Q14=\left (\mathbf{{{3-\sqrt{7}}\over 6}}, \mathbf{-{{7+2\sqrt {7}}\over
3}}, \mathbf{1}\right),
\nonumber \\
&& Q24=\left (\mathbf{{1\over 2}}, 
\mathbf{-{{2\sqrt{7}}\over {3}}}, \mathbf{2}\right ),~~~~~~~~~~  Q34=\left (\mathbf{1},
 \mathbf{-{{4\sqrt {7}}\over {3}}}, \mathbf{0}\right ).
\end{eqnarray}
The zero modes  include the third-family left-handed lepton doublet $L_3$,
one pair of Higgs doublets $H_u$ and $H_d$, and the right-handed
tau lepton $\tau^c$. From the trilinear term in the 7D bulk action, we obtain
the $\tau$ lepton Yukawa term
\begin{eqnarray}
 \int d^7 x \left[ \int d^2 \theta \ g_7 L_3 \tau^c H_d
 + h.c.\right].~\,
\end{eqnarray}

Thus, at the $M_{\rm GUT}$, we have
\begin{eqnarray}
g_1=g_2=g_3=y_{\tau}~,~\,
\end{eqnarray}
where  $y_{\tau}$ is the tau lepton Yukawa coupling.

\section{Split Supersymmetry}

The split SUSY proposal \cite{Arkani-Hamed:2004fb,Giudice:2004tc} abandons
light SUSY scalars as a solution to the gauge hierarchy problem. The MSSM scalars are
all assumed to be at an intermediate scale, except one Higgs doublet which is
fine-tuned to be light. The fermionic superpartners
remain light, preserving the gauge coupling unification and dark matter candidate.

For convenience, we assume the squarks, sleptons, charged and pseudoscalar Higgs to be
all degenerate at the scalar mass scale $m_S$. The particle content in the effective
theory beneath $m_S$ consists of the SM Higgs doublet H, as well as the Higgsinos
and gauginos. The Lagrangian is given by 

\begin{eqnarray}
{\cal L}&=&m^2 H^\dagger H-\frac{\lambda}{2}\left( H^\dagger H\right)^2
-\bigg[ h_u {\bar q} u\epsilon H^* 
+h_d {\bar q} d H
+h_e {\bar \ell} e H  \nonumber \\
&&+\frac{M_3}{2} {\tilde g}^A {\tilde g}^A
+\frac{M_2}{2} {\tilde W}^a {\tilde W}^a
+\frac{M_1}{2} {\tilde B} {\tilde B}
+\mu {\tilde H}_u^T\epsilon {\tilde H}_d \nonumber \\
&& +\frac{H^\dagger}{\sqrt{2}}\left( {\tilde g}_u \sigma^a {\tilde W}^a
+{\tilde g}_u^\prime {\tilde B} \right) {\tilde H}_u
+\frac{H^T\epsilon}{\sqrt{2}}\left(
-{\tilde g}_d \sigma^a {\tilde W}^a
+{\tilde g}_d^\prime {\tilde B} \right) {\tilde H}_d +{\rm h.c.}\bigg] ,
\label{lagr}
\end{eqnarray}
where $\epsilon =i\sigma_2$.

The SM Higgs doublet arises from a linear combination 
of the Higgs doublets $H_u$ and $H_d$ in the MSSM:
$H=-\cos\beta \epsilon H_d^*+\sin\beta H_u$.
By matching the Lagrangian in Eq. (\ref{lagr})
with the interaction terms of the Higgs doublets
$H_u$ and $H_d$ in the MSSM, the coupling constants of the
effective theory at the scale $m_S$ are obtained at tree level as follows:
\begin{eqnarray}
\lambda(m_S ) &=& \frac{\frac{3}{5}g_1^{2}(m_S )+ g_2^2(m_S )
}{4} \cos^22\beta , 
\nonumber\\
h_u(m_S )=y_u^*(m_S )\sin\beta , &&
h_{d,e}(m_S )=y_{d,e}^*(m_S )\cos\beta ,\nonumber\\
{\tilde g}_u (m_S )= g_2 (m_S )\sin\beta ,&&
{\tilde g}_d (m_S )= g_2 (m_S )\cos\beta ,\nonumber\\
{\tilde g}_u^\prime (m_S )= \sqrt{\frac{3}{5}}g_1(m_S ) \sin\beta ,&&
{\tilde g}_d^\prime (m_S )=  \sqrt{\frac{3}{5}}g_1(m_S )\cos\beta . 
\label{cond}
\end{eqnarray}

One expects to have threshold corrections to these relations from
integrating out the heavy scalars at the scale  $m_S$. 
But the mechanism which splits the scalar
 and  fermionic superpartners of the SM particles will
inevitably  suppress the $A$-terms, so,  $A\ll m_S$ and
there are no significant finite corrections from integrating out
the supersymmetric scalar particles \cite{Arkani-Hamed:2004fb}.

\section{Higgs Mass Predictions}

In this section, we will calculate the Higgs boson mass as a function of the scalar
mass scale $m_S$, assuming the gauge-top quark, gauge-bottom quark, 
and gauge-tau lepton Yukawa coupling unification.
Once $\tan\beta$ is fixed from gauge-Yukawa coupling unification,
the Higgs mass can be determined by running down all the couplings 
from the above boundary conditions~\footnote{Higgs boson
mass in split SUSY was calculated in Refs. \cite{Arkani-Hamed:2004fb,Giudice:2004tc,Arvanitaki:2004eu} 
for sample $\tan\beta$ values.}.

In our  numerical calculations,
we use two-loop renormalization group equation (RGE) running
 for the gauge couplings and one-loop RGE running for the Yukawa
and Higgs quartic couplings.  And the relevant RGEs are
given in the Appendix of Ref. \cite{Giudice:2004tc}. Also,
we use the $\overline{\rm MS}$ parameters:
the fine structure constant $\alpha^{-1}_{EM}(M_Z) = 127.918$ and the weak mixing
angle $\sin^2\theta_W(M_Z) = 0.23120$~\cite{pdg}, the bottom quark mass
$m_b(M_Z)=2.9$ GeV, and the top quark pole mass $M_{\rm
top}$(pole)$=172.5\pm2.3$ GeV~\cite{Beneke:2000hk}, where $M_Z$ is
the $Z$ boson mass.  We first run the parameters
from $M_Z$ up to $m_S$ and adjust them until the boundary conditions at $m_S$ 
in Eq. (\ref{cond}) are satisfied. The second step is running the gauge and Yukawa
couplings up to $M_{\rm GUT}$ with SUSY RGEs, and adjust $\tan\beta$ until the
relevant Yukawa coupling ($y_{t,b,\tau}$) equals to 
the gauge couplings at $M_{\rm GUT}$.

\begin{figure}[t]
\centering 
\includegraphics[angle=0, width=9cm]{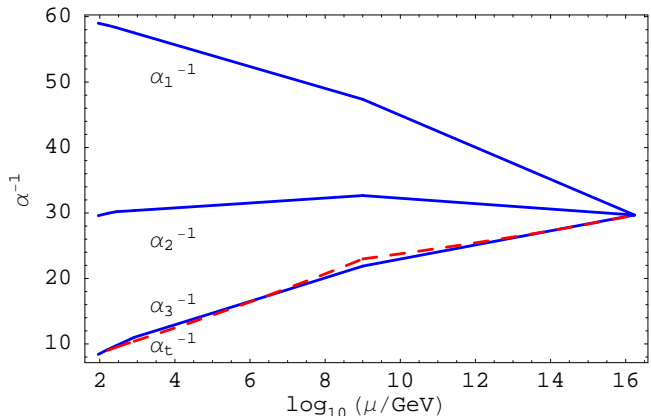} 
\vspace{-.5cm} 
\caption{Two-loop RGE evolution of gauge couplings (solid) and one-loop 
RGE evolution of top quark Yukawa
coupling (dashed), with $m_S=10^9$ GeV.} \label{ft}
\end{figure}

\begin{figure}[t]
 \centering
\includegraphics[angle=0, width=9cm]{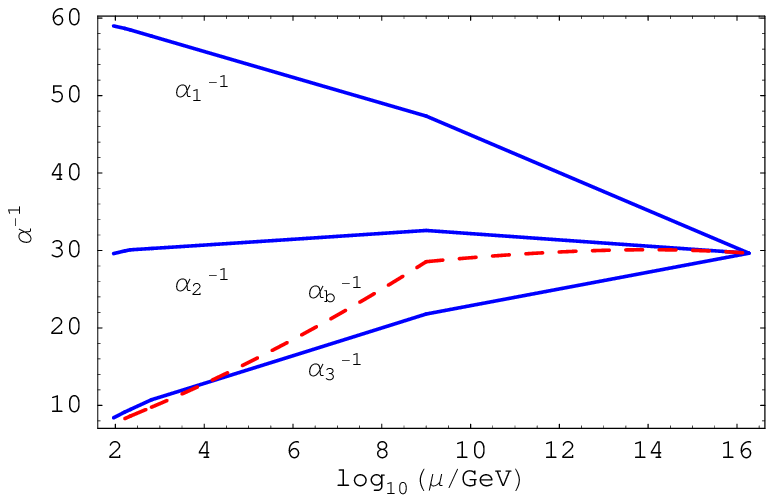} 
\vspace{-.5cm} 
\caption{Two-loop RGE evolution of gauge couplings (solid) and one-loop 
RGE evolution of bottom quark Yukawa
coupling (dashed), with $m_S=10^9$ GeV.} \label{fb}
\end{figure}

\begin{figure}[t]
 \centering
\includegraphics[angle=0, width=9cm]{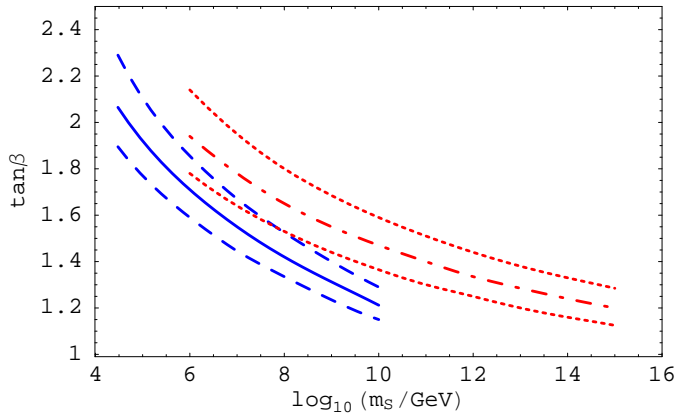} 
\vspace{-.5cm} 
\caption{$\tan\beta$ versus the scalar mass scale $m_S$ for  gauge-top quark
Yukawa coupling unification at $M_{\rm GUT}$.
The solid and dashed curves correspond to $M_3/M_{1/2}=3$, and
the dot-dashed and dotted curves to $M_3/M_{1/2}=10$
for $M_{\rm top}$(pole)$=172.5\pm2.3$ GeV.} \label{ftheta}
\end{figure}

\begin{figure}[t!]
 \centering
\includegraphics[angle=0, width=9cm]{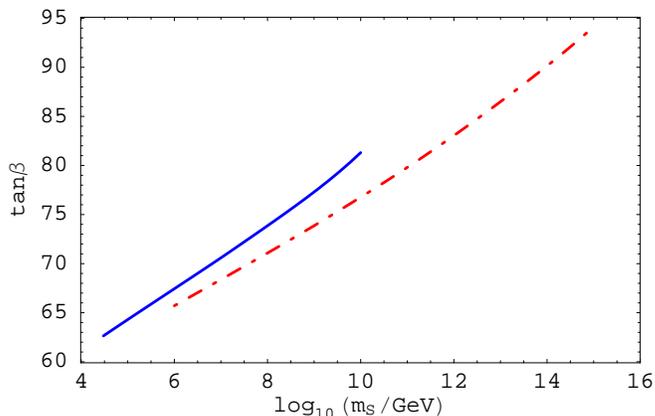} 
\vspace{-.5cm} 
\caption{$\tan\beta$ versus the scalar mass scale $m_S$ for gauge-bottom quark
Yukawa coupling unification at $M_{\rm GUT}$. The solid curve corresponds to 
$M_3/M_{1/2}=3$, and the dot-dashed curve to $M_3/M_{1/2}=10$. 
The uncertainties in bottom and top quark masses correspond to 5\% and 1\% 
uncertainties in $\tan\beta$, respectively.} \label{btheta}
\end{figure}

For gauge-top quark Yukawa coupling unification with $m_S=10^9$ GeV,
the SM gauge couplings (more precisely $\alpha_i^{-1}$) are plotted
in Fig. \ref{ft}, which also displays the top quark Yukawa coupling
$\alpha_t^{-1}\equiv 4\pi/y_t^2$. 
And Fig. \ref{fb} shows the couplings for 
gauge-bottom quark Yukawa coupling unification.
In addition, $\tan\beta$ as a function of $m_S$ for gauge-top and gauge-bottom quark Yukawa coupling
 unification is given in Figs.  \ref{ftheta} and \ref{btheta}, respectively. And 
$\tan\beta$ changes mildly with $m_S$ and remains $\sim55$ for gauge-tau
lepton Yukawa coupling unification.

\begin{figure}[t]
 \centering
\includegraphics[angle=0, width=9cm]{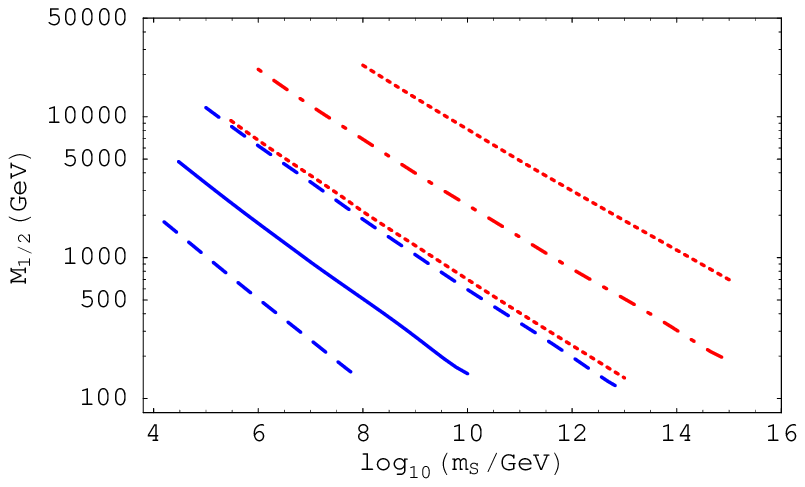} 
\vspace{-.5cm} 
\caption{$M_{1/2}$ versus $m_S$ for gauge-top quark
Yukawa coupling unification at $M_{\rm GUT}$.
 The solid and dashed curves correspond to 
$M_3/M_{1/2}=3$, and the dot-dashed and dotted curves to $M_3/M_{1/2}=10$
for $\alpha_s=0.119\pm0.003$.} \label{fgaugino}
\end{figure}

Usually, $\alpha_3(M_Z)$ is taken as a free parameter to be determined by gauge coupling unification.
In split SUSY, $\alpha_3(M_Z)$ decreases with increasing $m_S$, as expected.
However, the value of $\alpha_3(M_Z)$ depends on the gaugino masses as well as $m_S$,
decreasing as the gaugino masses increase \cite{Giudice:2004tc,Schuster:2004ad}. 
Let us denote $M_1,M_2,M_3$ as the bino, wino
and gluino masses, respectively.
For a simplified analysis, we assume $M_1=M_2=\mu=M_{1/2}$ where these masses
refer to the running masses at $M_{1/2}$, and compare two cases
with $M_3/M_{1/2}=3$ and $M_3/M_{1/2}=10$. 
The first case is typical for gaugino mass unification, while 
the second for anomaly mediated SUSY breaking \cite{Giudice:2004tc}. 
$\alpha_3(M_Z)$ is considerably higher for the latter case with fixed $M_{1/2}$ \cite{Dutta:2005zz}.
By fixing $\alpha_3(M_Z)=0.119\pm0.003$, we determine $M_{1/2}$ for a given $m_S$. 
Fig. \ref{fgaugino} shows $M_{1/2}$ for gauge-top quark Yukawa coupling unification, results
are similar for gauge-bottom quark or tau lepton Yukawa coupling  unification.

The dark matter requirements can be satisfied for $M_{1/2}$ ranging from a few
hundred GeV to a few TeV \cite{Giudice:2004tc,Pierce:2004mk}. It seems
possible to satisfy these requirements for any $m_S$ 
from about 1 TeV to $M_{\rm GUT}$.
However, the cosmological effects of a long lived gluino puts an upper bound on $m_S$ of
$10^9$ or $10^{11}$ GeV, where the latter value is for gluino mass less 
than 300 GeV \cite{Arvanitaki:2005fa}.
Together with gauge coupling unification
constraints, this disfavors the case $M_3/M_{1/2}=10$ in the absence of GUT-scale threshold
corrections. $m_S$ in the range $10^6$--$10^{11}$ GeV with $M_{1/2}\sim 1~{\rm TeV}$ 
and $M_3/M_{1/2}\sim3$ is qualitatively in good agreement with gauge coupling unification 
and dark matter abundance.

\begin{figure}[t!]
 \centering
\includegraphics[angle=0, width=9cm]{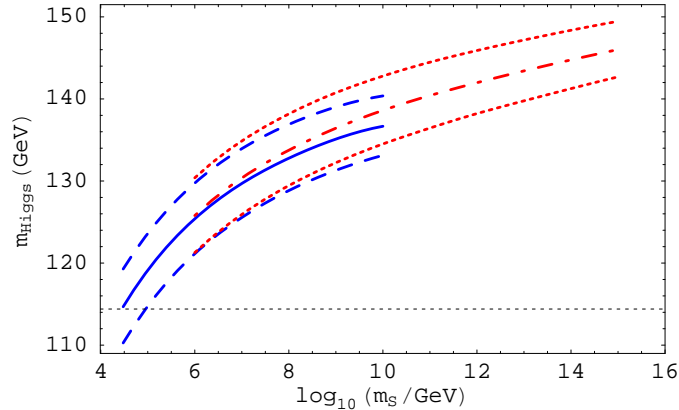} 
\vspace{-.5cm} 
\caption{Higgs boson mass $m_{\rm Higgs}$ versus the scalar mass scale $m_S$ for gauge-top quark
Yukawa coupling unification at $M_{\rm GUT}$.
The solid and dashed curves correspond to 
$M_3/M_{1/2}=3$, and the dot-dashed and dotted curves to $M_3/M_{1/2}=10$
for $M_{\rm top}$(pole)$=172.5\pm2.3$ GeV. } 
\label{fthiggs}
\end{figure}

\begin{figure}[t]
 \centering
\includegraphics[angle=0, width=9cm]{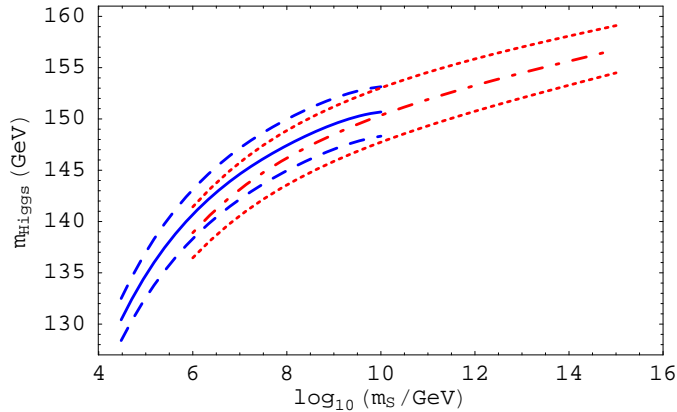} 
\vspace{-.5cm} 
\caption{Higgs boson mass $m_{\rm Higgs}$ versus the scalar mass scale $m_S$
 for gauge-bottom quark Yukawa coupling unification at $M_{\rm GUT}$.
The solid and dashed curves correspond to 
$M_3/M_{1/2}=3$, and the dot-dashed and dotted curves to $M_3/M_{1/2}=10$
for $M_{\rm top}$(pole)$=172.5\pm2.3$ GeV. } \label{fbhiggs}
\end{figure}

The Higgs mass is plotted in Figs. \ref{fthiggs} and \ref{fbhiggs} for
gauge-top and gauge-bottom quark Yukawa coupling unification, respectively (Higgs mass is 
practically identical for gauge-bottom quark and gauge-tau 
lepton Yukawa coupling unification.).
Although the main uncertainty in the Higgs boson mass prediction
 is due to the top quark mass uncertainty, varying
gaugino mass can also give a sizable effect. This effect is of
order 1 GeV or less for gauge-bottom quark Yukawa coupling unification. 
However, since the Higgs boson mass is
sensitive to $\tan\beta$ for small $\tan\beta$, the effect is larger (2-3 GeVs) for 
 gauge-top quark Yukawa coupling 
unification. The Higgs boson mass increases as $M_{1/2}$ increases, and decreases as
the ratio $M_3/M_{1/2}$ increases (for fixed $M_{1/2}$).

Because the Higgs boson mass is so sensitive to the top quark Yukawa coupling,
it is important to consider the one-loop corrections to the relation
between the running $\overline{\rm MS}$ Yukawa coupling and the pole
mass for the top quark. In the numerical calculations we use the 
results in Ref. \cite{Gray:1990yh}
(see also Ref. \cite{Schrempp:1996fb} for a review). For the relation
between the Higgs boson mass and Higgs quartic coupling, we use the results of
Ref. \cite{Sirlin:1985ux}, where we choose the renormalization
scale to be the top quark pole mass \cite{Hambye:1996wb}. We do not
consider the split SUSY corrections to these threshold effects. These
were considered in Ref. \cite{Binger:2004nn} and were found to be less important
than the SM corrections, generally affecting the Higgs mass $\lesssim2$ GeV.

\section{Conclusions}

We have constructed a class of 7D orbifold GUTs with ${\cal N}=1$
supersymmetry in which the SM gauge couplings and the top quark
(bottom quark or tau lepton) Yukawa coupling are unified at
the GUT scale. Assuming split supersymmetry, we can
reliably estimate the  SM Higgs boson mass.
 For gauge-top quark Yukawa coupling
 unification with $10^6$ GeV $\lesssim m_S\lesssim10^{11}$ GeV,
the Higgs boson mass range is $131\pm10$ GeV for $M_{\rm top}$(pole)$=172.5\pm2.3$ GeV.
For gauge-bottom quark (or tau lepton) 
Yukawa coupling unification, the Higgs boson mass range is $146\pm8$ GeV.

\section*{Acknowledgments}

This work is supported in part by DOE Grant   \# DE-FG02-84ER40163
(I.G.), \#DE-FG02-96ER40959 (T.L.),   \# DE-FG02-91ER40626 (Q.S. and
V.N.S.), and by a University of Delaware graduate fellowship
(V.N.S.).

\section*{Appendix A: Seven-Dimensional Orbifold Models}

We consider a 7D space-time $M^4\times T^2/Z_6 \times
S^1/Z_2$ with coordinates $x^{\mu}$, ($\mu = 0, 1, 2, 3$), $x^5$,
$x^6$ and $x^7$. The torus $T^2$ is homeomorphic to $S^1\times S^1$
and  the radii of the circles along the $x^5$, $x^6$ and $x^7$
directions are $R_1$, $R_2$, and $R'$, respectively. We define the
complex coordinate $z$ for $T^2$ and the real coordinate $y$ for
$S^1$,
\begin{eqnarray}
z \equiv{1\over 2} \left(x^5 + i x^6\right),~~~~~~~~~~ y \equiv x^7.
\end{eqnarray}
The torus $T^2$ can be defined by $C^1$ modulo the equivalent
classes:
\begin{eqnarray}
z \sim z+ \pi R_1 ,~~~~~~~~~~~ z \sim z +  \pi R_2 e^{{\rm
i}\theta}.
\end{eqnarray}
To obtain  the orbifold $T^2/Z_6$, we require that $R_1=R_2\equiv R$
and $\theta = \pi/3$. Then  $T^2/Z_6$  is obtained from $T^2$ by
moduloing the equivalent class
\begin{eqnarray}
\Gamma_T:~~~z \sim \omega  z,~\,
\end{eqnarray}
where $\omega =e^{{\rm i}\pi/3} $. There is one $Z_6$ fixed point
$z=0$, two $Z_3$ fixed points:
 $z=\pi R e^{{\rm i}\pi/6}/{\sqrt 3}$ and
$z=2 \pi R e^{{\rm i}\pi/6}/{\sqrt 3}$, and three $Z_2$ fixed
points: $z=\sqrt 3 \pi R e^{{\rm i}\pi/6}/2$, $z=\pi R/2$ and $z=
\pi R e^{{\rm i}\pi/3}/2$.
 The orbifold $S^1/Z_2$  is obtained from
$S^1$ by moduloing the equivalent class
\begin{eqnarray}
\Gamma_S:~~~y\sim -y~.~\,
\end{eqnarray}
There are two fixed points: $y=0$ and $y=\pi R'$.
The ${\cal N}=1$
supersymmetry in 7D has 16 supercharges corresponding to ${\cal
N}=4$ supersymmetry in 4D, and only the gauge multiplet can be
introduced in the bulk.  This multiplet can be decomposed under  4D
 ${\cal N}=1$ supersymmetry into a gauge vector
multiplet $V$ and three chiral multiplets $\Sigma_1$, $\Sigma_2$,
and $\Sigma_3$ in the adjoint representation, where the fifth and
sixth components of the gauge field, $A_5$ and $A_6$, are contained
in the lowest component of $\Sigma_1$, and the seventh component of
the gauge field $A_7$ is contained in the lowest component of
$\Sigma_2$.

We express the  bulk action in the Wess--Zumino gauge and 4D ${\cal
N}=1$ supersymmetry notation ~\cite{NMASWS}
\begin{eqnarray}
  {\cal S} &=& \int d^7 x \Biggl\{
  {\rm Tr} \Biggl[ \int d^2\theta \left( \frac{1}{4 k g^2}
  {\cal W}^\alpha {\cal W}_\alpha + \frac{1}{k g^2}
  \left( \Sigma_3 \partial_z \Sigma_2 + \Sigma_1 \partial_y \Sigma_3
   - \frac{1}{\sqrt{2}} \Sigma_1
  [\Sigma_2, \Sigma_3] \right) \right)
\nonumber\\
  &&
+ h.c. \Biggr]
 + \int d^4\theta \frac{1}{k g^2} {\rm Tr} \Biggl[
  (\sqrt{2} \partial_z^\dagger + \Sigma_1^\dagger) e^{-V}
  (-\sqrt{2} \partial_z + \Sigma_1) e^{V}
 + \partial_z^\dagger e^{-V} \partial_z e^{V}
\nonumber\\
  && +
  (\sqrt{2} \partial_y + \Sigma_2^\dagger) e^{-V}
  (-\sqrt{2} \partial_y + \Sigma_2) e^{V}
 + \partial_y e^{-V} \partial_y e^{V}
+ {\Sigma_3}^\dagger e^{-V} \Sigma_3 e^{V} \Biggr] \Biggr\},~\,
\label{action7}
\end{eqnarray}
where $k$ is the normalization of the group generator,  and ~${\cal
W_{\alpha}}$~ denotes the gauge field strength.  From the above
action, we obtain the transformations of the vector multiplet:
\begin{eqnarray}
  V(x^{\mu}, ~\omega z, ~\omega^{-1} {\bar z},~y) &=& R_{\Gamma_T}
 V(x^{\mu}, ~z, ~{\bar z},~y) R_{\Gamma_T}^{-1}~,~\,
\label{TVtrans}
\end{eqnarray}
\begin{eqnarray}
  \Sigma_1(x^{\mu}, ~\omega z, ~\omega^{-1} {\bar z},~y) &=&
\omega^{-1} R_{\Gamma_T} \Sigma_1(x^{\mu}, ~z, ~{\bar z},~y)
R_{\Gamma_T}^{-1}~,~\, \label{T1trans}
\end{eqnarray}
\begin{eqnarray}
   \Sigma_2(x^{\mu}, ~\omega z, ~\omega^{-1} {\bar z},~y) &=&
 R_{\Gamma_T}
\Sigma_2(x^{\mu}, ~z, ~{\bar z},~y)  R_{\Gamma_T}^{-1}~,~\,
\label{T2trans}
\end{eqnarray}
\begin{eqnarray}
 \Sigma_3(x^{\mu}, ~\omega z, ~\omega^{-1} {\bar z},~y)  &=&
\omega R_{\Gamma_T} \Sigma_3(x^{\mu}, ~z, ~{\bar z},~y)
R_{\Gamma_T}^{-1}~,~\, \label{T3trans}
\end{eqnarray}
\begin{eqnarray}
  V(x^{\mu}, ~z, ~ {\bar z},~-y) &=& R_{\Gamma_S}
 V(x^{\mu}, ~z, ~{\bar z},~y) R_{\Gamma_S}^{-1}~,~\,
\label{SVtrans}
\end{eqnarray}
\begin{eqnarray}
  \Sigma_1(x^{\mu}, ~ z, ~ {\bar z},~-y) &=&
 R_{\Gamma_S}
\Sigma_1(x^{\mu}, ~z, ~{\bar z},~y) R_{\Gamma_S}^{-1}~,~\,
\label{S1trans}
\end{eqnarray}
\begin{eqnarray}
   \Sigma_2(x^{\mu}, ~ z, ~ {\bar z},~-y) &=&
-  R_{\Gamma_S} \Sigma_2(x^{\mu}, ~z, ~{\bar z},~y)
R_{\Gamma_S}^{-1}~,~\, \label{S2trans}
\end{eqnarray}
\begin{eqnarray}
 \Sigma_3(x^{\mu}, ~ z, ~ {\bar z},~-y)  &=&
- R_{\Gamma_S} \Sigma_3(x^{\mu}, ~z, ~{\bar z},~y)
R_{\Gamma_S}^{-1}~,~\, \label{S3trans}
\end{eqnarray}
where we introduce  non-trivial transformation $R_{\Gamma_T}$ and
$R_{\Gamma_S}$ to break the bulk gauge group $G$.

The $Z_6\times Z_2$ transformation properties for the decomposed
components of $V$, $\Sigma_1$, $\Sigma_2$,  and $\Sigma_3$
in our $SU(7)$ models are given by
\begin{equation}
V : \left(
\begin{array}{cccc}
(1, +) & (\omega^{-n_1}, +) & (\omega^{-n_1}, -) &
(\omega^{-n_2}, -)  \\
(\omega^{n_1}, +) & (1, +) & (1, -) & (\omega^{n_1-n_2}, -) \\
(\omega^{n_1}, -) & (1, -) & (1, +) & (\omega^{n_1-n_2}, +) \\
(\omega^{n_2}, -) & (\omega^{n_2-n_1}, -) & (\omega^{n_2-n_1}, +) &
(1, +)
\end{array}
\right)  +  (1, +) ~,~\, \label{bb7}
\end{equation}
\begin{equation}
\Sigma_1 : \left(
\begin{array}{cccc}
(\omega^{-1}, +) & (\omega^{-n_1-1}, +) & (\omega^{-n_1-1}, -) &
(\omega^{-n_2-1}, -)  \\
(\omega^{n_1-1}, +) & (\omega^{-1}, +) & (\omega^{-1}, -) & (\omega^{n_1-n_2-1}, -) \\
(\omega^{n_1-1}, -) & (\omega^{-1}, -) & (\omega^{-1}, +) & (\omega^{n_1-n_2-1}, +) \\
(\omega^{n_2-1}, -) & (\omega^{n_2-n_1-1}, -) & (\omega^{n_2-n_1-1},
+)  & (\omega^{-1}, +)
\end{array}
\right) +  (\omega^{-1}, +) ~,~\, \label{bb8}
\end{equation}
\begin{equation}
\Sigma_2 : \left(
\begin{array}{ccccc}
(1, -) & (\omega^{-n_1}, -) & (\omega^{-n_1}, +) &
(\omega^{-n_2}, +) \\
(\omega^{n_1}, -) & (1, -) & (1, +) & (\omega^{n_1-n_2}, +) \\
(\omega^{n_1}, +) & (1, +) & (1, -) & (\omega^{n_1-n_2}, -) \\
(\omega^{n_2}, +) & (\omega^{n_2-n_1}, +) & (\omega^{n_2-n_1}, -) &
(1, -)
\end{array}
\right) +  (1, -) ~,~\, \label{bb9}
\end{equation}
\begin{equation}
\Sigma_3 : \left(
\begin{array}{ccccc}
(\omega, -) & (\omega^{-n_1+1}, -) & (\omega^{-n_1+1}, +) &
(\omega^{-n_2+1}, +)  \\
(\omega^{n_1+1}, -) & (\omega, -) & (\omega, +) & (\omega^{n_1-n_2+1}, +) \\
(\omega^{n_1+1}, +) & (\omega, +) & (\omega, -) & (\omega^{n_1-n_2+1}, -) \\
(\omega^{n_2+1}, +) & (\omega^{n_2-n_1+1}, +) & (\omega^{n_2-n_1+1},
-) & (\omega, -)
\end{array}
\right) +  (\omega, -) ~,~\, \label{bb10}
\end{equation}
where the zero modes transform as $(1,+)$.

From Eqs. (\ref{bb7})--(\ref{bb10}), we find that  the 7D ${\cal N}
= 1 $ supersymmetric gauge symmetry $SU(7)$  is broken down to  4D
${\cal N}=1$ supersymmetric gauge symmetry $SU(3)_C\times
SU(2)_L\times U(1)_Y \times U(1)_{\alpha} \times U(1)_{\beta} $
 \cite{Li:2001tx}. In addition, there are  zero modes from the chiral
multiplets $\Sigma_1$, $\Sigma_2$ and $\Sigma_3$ which play an
important role in gauge--Yukawa coupling unification \cite{GY}.

 \clearpage

\end{document}